\documentclass{PoS}

\title{Nonperturbative tests of the renormalization of mixed clover-staggered currents in lattice QCD}

\ShortTitle{Nonperturbative tests of clover-staggered currents}

\author{\speaker{Bipasha Chakraborty}\\SUPA, School of Physics and Astronomy, University of Glasgow, Glasgow G12 8QQ, UK\\
        E-mail: \email{b.chakraborty.1@research.gla.ac.uk}}
\author{Christine Davies\\SUPA, School of Physics and Astronomy, University of Glasgow, Glasgow G12 8QQ, UK\\
         E-mail: \email{Christine.Davies@glasgow.ac.uk}}
\author{Gordon Donald\\School of Mathematics, Trinity College, Dublin 2, Ireland}
\author{Rachel Dowdall\\DAMTP, University of Cambridge, Wilberforce Road, Cambridge CB3 0WA, UK}
\author{Jonna Koponen\\SUPA, School of Physics and Astronomy, University of Glasgow, Glasgow G12 8QQ, UK}
\author{G. Peter Lepage\\Laboratory of Elementary Particle Physics, Cornell University, Ithaca, NY 14853, USA}
\author{HPQCD Collaboration}

\abstract{The Fermilab Lattice and MILC collaborations have shown in one-loop lattice QCD perturbation theory that the renormalization constants of vector and axial-vector mixed clover-asqtad currents are closely related to the product of those for clover-clover and asqtad-asqtad (local) vector currents. To be useful for future higher precision calculations this relationship must be valid beyond one-loop and very general. We test its validity nonperturbatively using clover and Highly Improved Staggered (HISQ) strange quarks, utilising the absolute normalization of the HISQ temporal axial current. We find that the renormalization of the mixed current differs from the square root of the product of the pure HISQ and pure clover currents by 2-3\%. We also compare discretization errors between the clover and HISQ formalisms.}

\FullConference{31st International Symposium on Lattice Field Theory LATTICE 2013\\
                 July 29-August 3, 2013\\
                 Mainz, Germany}

\begin{document}

\section{Motivation}
\begin{figure}[h]
\centering
   \includegraphics[width=7.5cm]{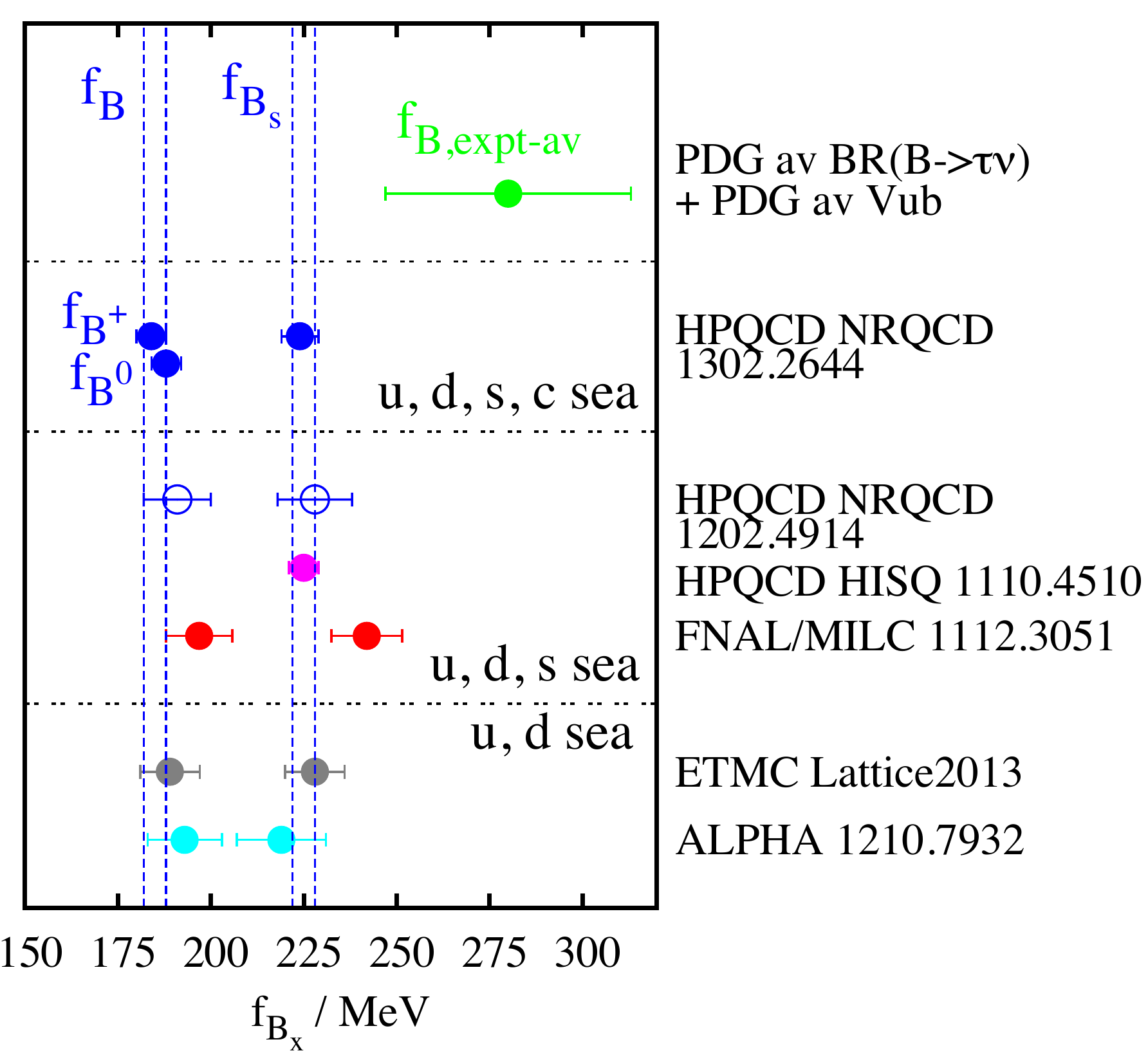}
   \caption{Results for the $B$ and $B_s$ meson decay constants using different formalisms and number of flavours of sea quarks.}
   \label{fig:B_Bs}
\end{figure}

The leptonic decay constants of the heavy-light mesons are important physical quantities in lattice QCD. They are needed to extract CKM matrix elements~\cite{Davies} from the measurements of the decay rate and thus need to be calculated precisely from lattice QCD to test the unitarity of the CKM matrix in the Standard Model as stringently as possible. A good test of lattice QCD is the comparison of results using different quark formalisms. Results for the $B$ and $B_s$ meson decay constants are summarized in Fig.~\ref{fig:B_Bs} and the tension with experiment for $f_B$ shown~\cite{PDG}. The HPQCD collaboration have obtained separate results using NRQCD $b$ quarks~\cite{RachelfB} and HISQ $b$ quarks~\cite{McNeile}, both combined with HISQ light quarks, which are consistent within their ~2\% errors for $f_{B_s}$. The systematic uncertainties for the two methods are very different, with the NRQCD $b$ quark calculation having a significant contribution from the uncertainty in the current renormalization. There is no such uncertainty in the HISQ $b$ quark calculation but there the error is dominated by discretization errors. An alternative method from the  Fermilab lattice/MILC collaborations uses heavy clover $b$ quarks. This method also gives a reasonably consistent result but with much larger errors coming from a number of sources. Here we test the method for determining the current renormalization in this formalism to see if the estimate of the error from that source is robust.
\section{Our method}
\begin{figure}[h]                                                                     
\centering
   \includegraphics[width=7.5cm]{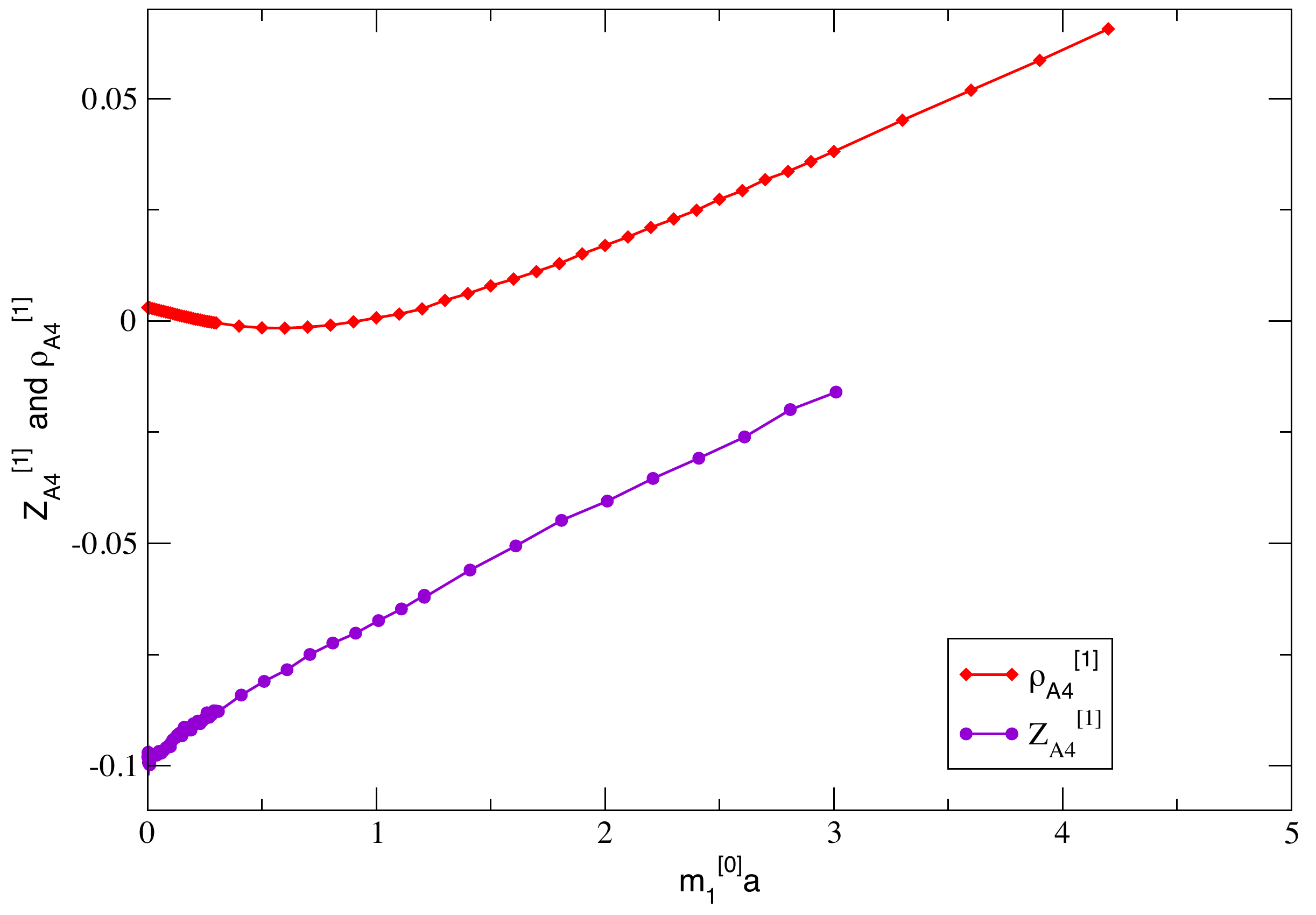}
   \caption{The one-loop coefficients of $\rho_{A^4_{Qq}}$ and $Z_{A^4_{Qq}}$ for the clover-asqtad case plotted against the bare clover quark mass~\cite{El-Khadra}.}
   \label{fig:El-Khadra}                                                          
\end{figure}

The Fermilab/MILC collaboration used asqtad light valence quarks and clover $b$ and $c$ quarks for calculating the B and D meson decay constants~\cite{BD}. In doing so, the heavy-light axial current renormalization constant $Z_{A^{\mu}_{Qq}}$ was calculated partly non-perturbatively and partly in one-loop perturbation theory. They proposed a definition relating the renormalization constants for the heavy-light temporal axial current and the square roots of the renormalization constants for the heavy-heavy and light-light local temporal vector currents given by
\begin{eqnarray}
Z_{A^4_{Qq}} = \rho_{A^4_{Qq}} \sqrt{Z_{V^4_{qq}} Z_{V^4_{QQ}}}   .\nonumber 
\end{eqnarray}
It has been shown in one-loop perturbation theory that the correction factor $\rho_{A^4_{Qq}}$ has a small one-loop coefficient (see Fig.~\ref{fig:El-Khadra} from~\cite{El-Khadra}). This result is used to argue that $\rho_{A^4_{Qq}}$ is close to unity for higher orders too, although this claim has not been explicitly demonstrated. We have carried out a completely non-perturbative calculation of the Z factors to test the validity of the Fermilab/MILC claim beyond one-loop. We have used HISQ-HISQ, clover-clover and mixed clover-HISQ currents made of strange valence quarks to see how close the $\rho_{A^4_{Qq}}$ is to unity. Since the claim must be general to be useful, it is not necessary for us to use asqtad quarks to test it. Instead we use HISQ quarks since they are further improved~\cite{Eduardo} and are being used in current state-of-the-art calculations~\cite{Rachel2}.


We have chosen the strange quark as the valence quark as its mass too falls within the light quark mass region where (in fact, see Fig.~\ref{fig:El-Khadra}) the claim for the smallness of the one-loop coefficient of $\rho_{Qq}$ holds~\cite{El-Khadra}. The valence strange quark mass was tuned to give the mass of a fictitious pseudoscalar meson known as $\eta_s$ to be 688.5 MeV~\cite{DowdallVus}. We made the $\eta_s$ from two HISQ strange quarks, two clover strange quarks and mixing one HISQ and one clover strange quark. 

The decay constant of the $\eta_s$ is related to the matrix element of the temporal axial current between the $\eta_s$ and the vacuum and is given by $\langle 0|A^4|\eta_s(0)\rangle = m_{\eta_s}f_{\eta_s}$ when the $\eta_s$ is at rest. We have utilized the fact that the HISQ-HISQ temporal axial current is absolutely normalized and via the PCAC relation we can then obtain $f_{\eta_s}$ with no normalization factor. We then obtained the temporal axial vector current renormalization for the clover-HISQ discretization using the relation $Z_{A^4_{Cl-H}}f_{\eta_s}^{Cl-H}=f_{\eta_s}^{H-H}$. 

The renormalization factors for the clover-clover and HISQ-HISQ local vector currents are obtained by demanding at zero momentum transfer that $1 = Z_{V^4_{qq}} \langle H_q|V^4_{qq}|H_q\rangle$. The matrix element is extracted by fitting the two-point and three-point correlators simultaneously. A point to note: We need to use an unstaggered (clover in our case) quark propagator as the spectator quark in the three point correlators for the HISQ-HISQ local vector current since a purely staggered three-point function with the same mesons at either end would vanish in this case (because the local vector current is not a taste-singlet). 


For our calculation, we have used three lattice ensembles with lattice spacings $a\approx$0.15fm (very coarse), 0.12fm (coarse) and 0.09fm (fine). The ensembles were generated by MILC using the light, strange and charm HISQ quarks in the sea where $m_l/m_s\approx 0.2$~\cite{MILCensemble}. The valence HISQ strange quark mass was tuned using the Wilson flow parameter $w_0$ to set the lattice spacing~\cite{DowdallVus} and listed in table~\ref{table:simpar}. There we have also listed tuned valence clover ${\kappa}_s$ on each ensemble, the number of configurations used $n_{cfg}$ and the number of time sources per configuration $n_t$.
\begin{table}[h]
\caption{List of simulation parameters}
\centering
\begin{tabular}{cccccc}
\hline
\hline
Set  & $a$(fm) & ${am_s}^{HISQ,val}$ & ${\kappa_s}^{Clover,val}$  & $n_{cfg}$ & $n_t$  \\
\hline
Very Coarse &  0.1543(8) & 0.0705 & 0.14082 & 1021 & 12  \\
\hline
Coarse &  0.1241(7) & 0.0541 & 0.13990 & 527 & 16  \\
\hline
Fine &  0.0907(5) & 0.0376 & 0.13862 & 504 & 16  \\
\hline
\hline
\end{tabular}
\label{table:simpar}
\end{table}


\section{Our Results}

Table~\ref{table:results} summarizes our results for testing the relation $Z_{{A^4}_{Cl-H}} = \rho_{{A^4}_{Cl-H}}\sqrt{Z_{{V^4}_{Cl-Cl}}Z_{{V^4}_{H-H}}}$.
\begin{table}[tbh]
\caption{Our results for $Z_{A^4}$, $Z_{V^4}$ and $\rho_{{A^4}_{ab}}$}
\centering
\begin{tabular}{ccccc}
\hline
\hline
Set &  Combinations  &  $Z_{A^4}$  &  $Z_{V^4}$  &   $\rho_{{A^4}_{ab}}$ \\
\hline
VC & H-H & 1.000 & 0.9887(20) & -  \\
  & Cl-Cl & 0.2046(4) & 0.2045(3) & - \\
  & Cl-H & 0.4642(6) & - & 1.0322(21) \\
\hline
C &  H-H &  1.000 & 0.9938(17) & - \\
  & Cl-Cl & 0.2096(4) & 0.2071(4) & -\\
  & Cl-H &  0.4656(4) & - & 1.0263(36) \\
\hline
F &  H-H &  1.000  & 0.9944(10) & -\\
  & Cl-Cl &  0.2152(4) & 0.2116(4) & -\\
  & Cl-H &  0.4679(7)  &  -  & 1.0199(33) \\
\hline
\hline
\end{tabular}
\label{table:results}
\end{table}

\begin{figure}[h]
\centering
\includegraphics[width=8.0cm]{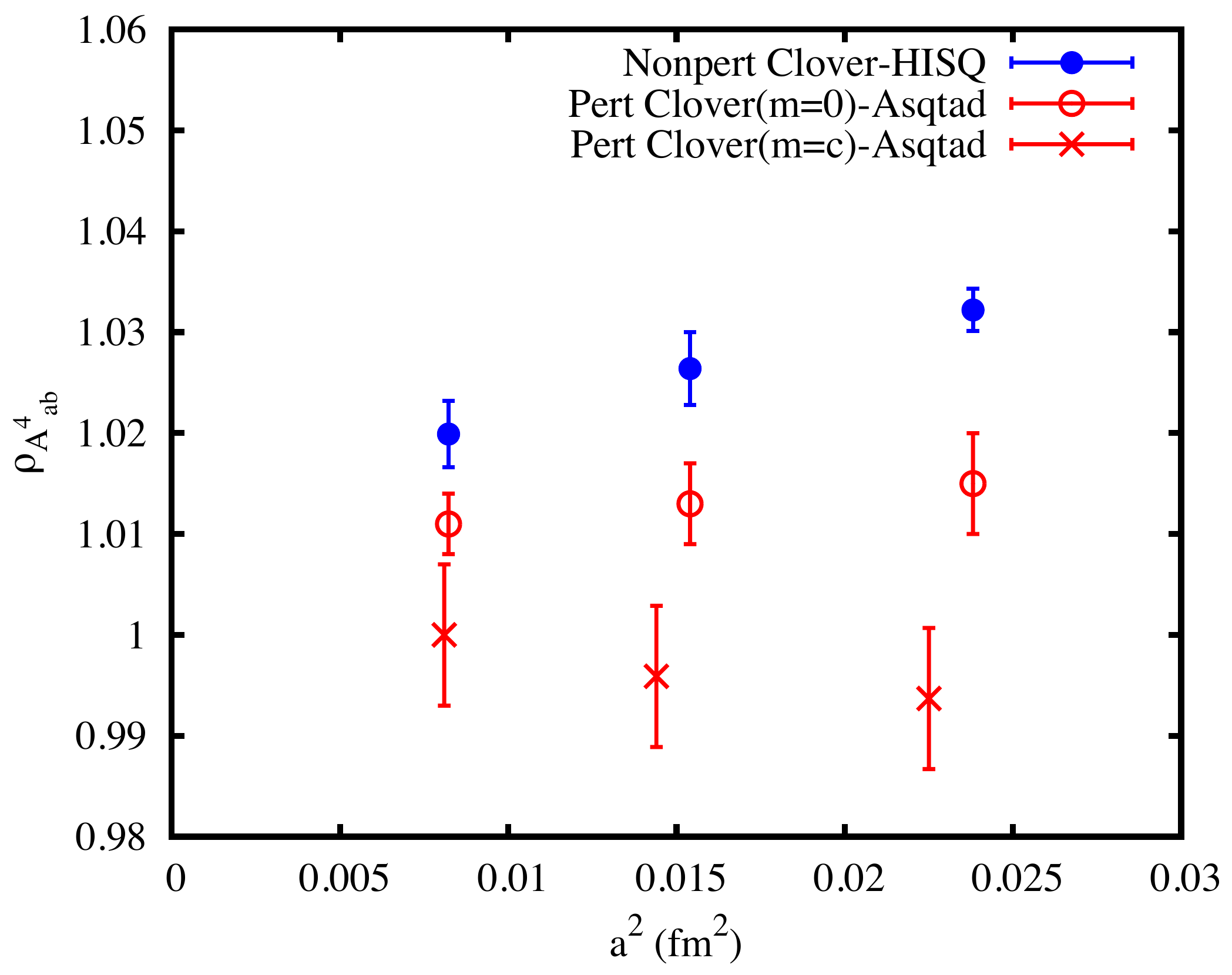}
\caption{Our results for $\rho_{{A^4}_{ab}}$ (dots), plotted against the square of the lattice spacing and compared to the one-loop results from Fermilab/MILC for mixed clover-asqtad currents with clover charm (crosses) and clover light (open circles) quarks.}

\label{fig:our_results}
\end{figure}

After calculating the Z factors completely nonperturbatively we find $\rho_{{A^4}_{Cl-H}}$ is indeed close to 1.0 up to all orders of lattice perturbation theory with a maximum deviation of $\sim3\%$ on very coarse lattice. Our results for $\rho_{{A^4}_{Cl-H}}$ are shown and compared to the Fermilab/MILC results for clover-asqtad using their one-loop perturbative coefficients and two different values of the clover quark mass in Fig.~\ref{fig:our_results}.
The equivalent perturbative results for clover-HISQ could be different because asqtad and HISQ are different formalisms. However, the fact that our numbers do lie further from 1 than the Fermilab/MILC results suggests that it may be sensible to increase the perturbative errors in their results to $2\%$. This would then allow for the possibility that the clover-asqtad all-orders result is as far from 1 as our clover-HISQ result. 

We also note from the table~\ref{table:results} that $Z_{{V^4}_{Cl-Cl}}$ and $Z_{{A^4}_{Cl-Cl}}$ are very close to each other, despite the fact that clover quarks break chiral symmetry.

\section{Discretization effects}
\begin{figure}[h]
\centering
\includegraphics[width= 8.0cm]{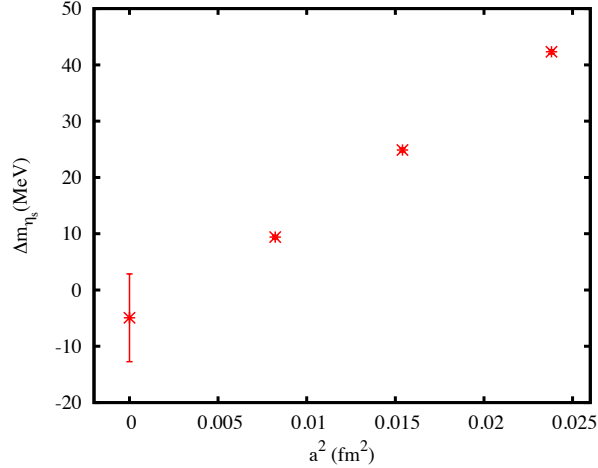}
\caption{The difference of the $\eta_s$ mass obtained from the HISQ-HISQ and clover-HISQ methods}
\label{fig:massdiff}
\end{figure}

We also discuss the discretization effects coming from the different methods used in our present calculation. The difference of the $\eta_s$ mass obtained from the HISQ-HISQ and clover-HISQ correlators (when both HISQ and clover quark masses are tuned to that of the strange quark) is plotted in Fig.~\ref{fig:massdiff} and is clearly a discretization effect, vanishing quadratically as the lattice spacing goes to zero.

\begin{figure}[h]
\centering
\includegraphics[width=7.2cm]{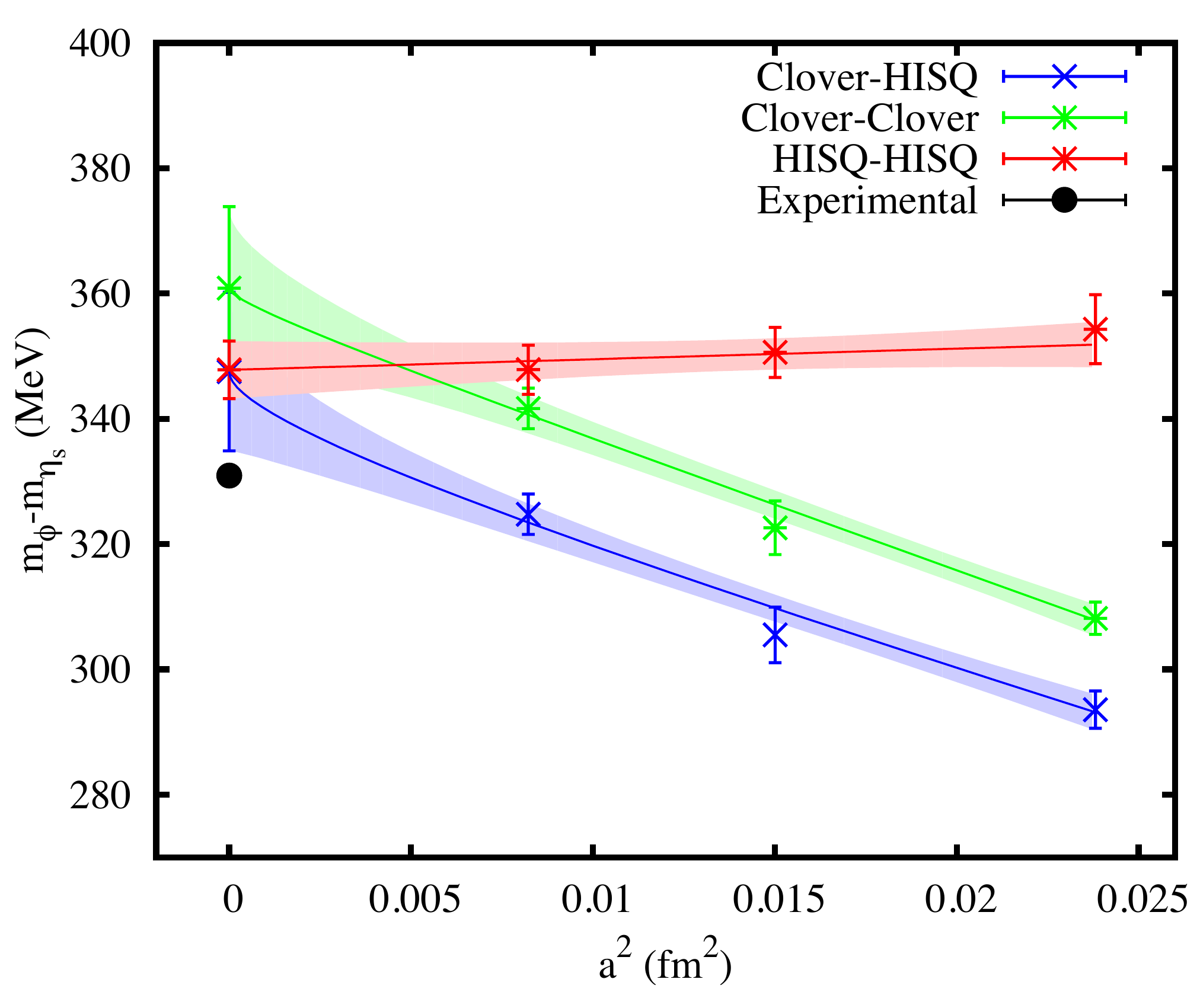}\hspace{0.2em}\includegraphics[width=7.4cm]{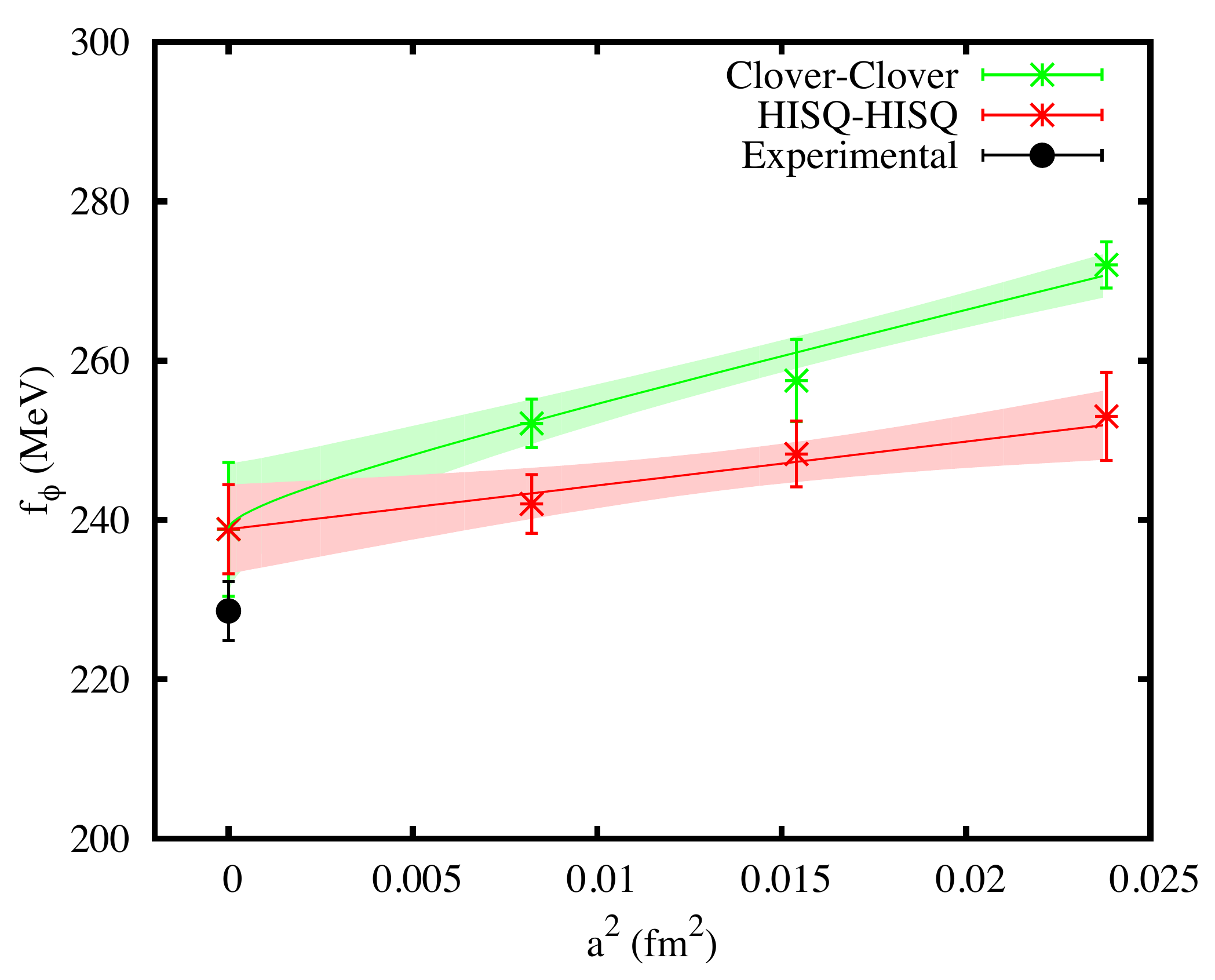}
\caption{On the left: $m_{\phi}-m_{{\eta}_s}$ calculated with different quark formalisms and extrapolated to $a=0$. On the right $f_{\phi}$ calculated with different quark formalisms and extrapolated to $a=0$.}
\label{fig:massextrap}
\end{figure}

For another comparison of discretization effects we calculate the mass of the vector meson $\phi$ and then extrapolate $m_{\phi}-m_{{\eta}_s}$ to the continuum $a=0$. From the extrapolation plot in Fig.~\ref{fig:massextrap} we see all three methods of calculating correlators agree in the continuum limit as expected. The HISQ-HISQ discretization errors are much smaller than the clover-clover and the clover-HISQ discretizations. The accurate HISQ-HISQ results show a value in continuum limit which is higher than experiment. The $\phi$ is not a gold-plated meson, having a strong decay to $K\overline{K}$ and further study is needed to uncover what the impact of this decay channel is.

We also calculate the $\phi$ decay contant using the renomalization constants $Z_{{V^4}_{Cl-Cl}}$ and $Z_{{V^4}_{H-H}}$ determined earlier. This is also plotted as a function of lattice spacing in Fig.~\ref{fig:massextrap} with extrapolation to the continuum. As before, we find that all discretizations agree in the continuum and again the HISQ-HISQ case shows the smallest discretization errors. In this case, the lattice $f_{\phi}$ results match with the experimental result obtained from $\Gamma(\phi \rightarrow e^+ e^-)$~\cite{PDG} within $1.5\sigma$.



\section{Conclusion and Ongoing Work}
The following conclusions are drawn based on our present work:
\begin{itemize}
\item $\rho_{{A^4}_{Cl-H}}$ is close to 1.0 with a maximum $\sim3\%$ deviation using the nonperturbative lattice calculations with pure HISQ, pure clover and mixed clover-HISQ currents.
\item Based on our numbers we would recommend increasing the errors in the Fermilab/MILC results from $0.7\%$ to $2.0\%$ in one loop perturbative calculation of $\rho_{{A^4}_{Cl-H}}$.
\item Discretization errors are significantly smaller for HISQ than clover.
\end{itemize}
We are also studying $\rho_{{V^4}_{Cl-H}}$ using $Z_{{V^4}_{Cl-H}}$. We will extend our study of the vector meson $\phi$ to the MILC gluon field configurations that include physical u/d sea quarks to test more carefully the effect of it not being gold-plated.

\subsection*{\bf{Acknowledgements}} 

We are grateful to the MILC collaboration for the use of their configurations and to A. Kronfeld for useful discussions. 
Computing was done on the Darwin supercomputer at the University of Cambridge as part of STFC's DiRAC facility. We are grateful to the Darwin support staff for assistance. Funding for this work came from the NSF, the Royal Society, the Wolfson Foundation and STFC.



\bibliographystyle{h-physrev.bst}
\bibliography{sources}{}

\begin{thebibliography}{10}

\bibitem{Davies}
C.~Davies,
\newblock PoS {\bf LATTICE2011}, 019 (2011), 1203.3862.

\bibitem{PDG}
Particle Data Group, J.~Beringer {\em et~al.},
\newblock Phys. Rev. D {\bf 86}, 010001 (2012).

\bibitem{RachelfB}
HPQCD Collaboration, R.~Dowdall {\em et~al.},
\newblock (2013), 1309.4610.

\bibitem{McNeile}
C.~McNeile, C.~Davies, E.~Follana, K.~Hornbostel, and G.~Lepage,
\newblock Phys.Rev. {\bf D85}, 031503 (2012), 1110.4510.

\bibitem{BD}
Fermilab Lattice and MILC Collaborations, A.~Bazavov {\em et~al.},
\newblock Phys. Rev. D {\bf 85}, 114506 (2012).

\bibitem{El-Khadra}
A.~X. El-Khadra, E.~Gamiz, A.~S. Kronfeld, and M.~A. Nobes,
\newblock PoS {\bf LAT2007}, 242 (2007), 0710.1437.

\bibitem{Eduardo}
HPQCD Collaboration, E.~Follana {\em et~al.},
\newblock Phys. Rev. D {\bf 75}, 054502 (2007).

\bibitem{Rachel2}
HPQCD Collaboration, R.~Dowdall, C.~Davies, R.~Horgan, C.~Monahan, and
  J.~Shigemitsu,
\newblock Phys.Rev.Lett. {\bf 110}, 222003 (2013), 1302.2644.

\bibitem{DowdallVus}
HPQCD Collaboration, R.~Dowdall, C.~Davies, G.~Lepage, and C.~McNeile,
\newblock Phys.Rev. {\bf D88}, 074504 (2013), 1303.1670.

\bibitem{MILCensemble}
MILC Collaboration, A.~Bazavov {\em et~al.},
\newblock Phys. Rev. D {\bf 87}, 054505 (2013).

\end{thebibliography}


@article{upsilon,
  title = {The Upsilon spectrum and the determination of the lattice spacing from lattice QCD including charm quarks in the sea},
  author = {Dowdall, R. J. and Colquhoun, B. and Daldrop, J. O. and Davies, C. T. H. and Kendall, I. D. and Follana, E. and Hammant, T. C. and Horgan, R. R. and Lepage, G. P. and Monahan, C. J. and M\"uller, E. H.},
  collaboration = {HPQCD Collaboration},
  journal = {Phys. Rev. D},
  volume = {85},
  issue = {5},
  pages = {054509},
  numpages = {38},
  year = {2012},
  month = {Mar},
  doi = {10.1103/PhysRevD.85.054509},
  url = {http://link.aps.org/doi/10.1103/PhysRevD.85.054509},
  publisher = {American Physical Society}
}


@article{BD,
  title = {$B$- and $D$-meson decay constants from three-flavor lattice QCD},
  author = {Bazavov, A. and Bernard, C. and Bouchard, C. M. and DeTar, C. and Di Pierro, M. and El-Khadra, A. X. and Evans, R. T. and Freeland, E. D. and G\'amiz, E. and Gottlieb, Steven and Heller, U. M. and Hetrick, J. E. and Jain, R. and Kronfeld, A. S. and Laiho, J. and Levkova, L. and Mackenzie, P. B. and Neil, E. T. and Oktay, M. B. and Simone, J. N. and Sugar, R. and Toussaint, D. and Van de Water, R. S.},
  collaboration = {Fermilab Lattice and MILC Collaborations},
  journal = {Phys. Rev. D},
  volume = {85},
  issue = {11},
  pages = {114506},
  numpages = {37},
  year = {2012},
  month = {Jun},
  doi = {10.1103/PhysRevD.85.114506},
  url = {http://link.aps.org/doi/10.1103/PhysRevD.85.114506},
  publisher = {American Physical Society}
}


@article{Lepagenotes,
title = "Constrained curve fitting ",
journal = "Nuclear Physics B - Proceedings Supplements ",
volume = "106–107",
number = "0",
pages = "12 - 20",
year = "2002",
note = "<ce:title>LATTICE 2001 Proceedings of the \{XIXth\} International Symposium on Lattice Field Theory</ce:title> ",
issn = "0920-5632",
doi = "10.1016/S0920-5632(01)01638-3",
url = "http://www.sciencedirect.com/science/article/pii/S0920563201016383",
author = "G.P. Lepage and B. Clark and C.T.H. Davies and K. Hornbostel and P.B. Mackenzie and C. Morningstar and H. Trottier"
}


@article{stag,
  title = {Highly improved staggered quarks on the lattice with applications to charm physics},
  author = {Follana, E. and Mason, Q. and Davies, C. and Hornbostel, K. and Lepage, G. P. and Shigemitsu, J. and Trottier, H. and Wong, K.},
  collaboration = {HPQCD Collaboration},
  journal = {Phys. Rev. D},
  volume = {75},
  issue = {5},
  pages = {054502},
  numpages = {23},
  year = {2007},
  month = {Mar},
  doi = {10.1103/PhysRevD.75.054502},
  url = {http://link.aps.org/doi/10.1103/PhysRevD.75.054502},
  publisher = {American Physical Society}
}


@article{PhysRevD.81.034506,
  title = {Precise determination of the lattice spacing in full lattice QCD},
  author = {Davies, C. T. H. and Follana, E. and Kendall, I. D. and Lepage, G. Peter and McNeile, C.},
  collaboration = {HPQCD Collaboration},
  journal = {Phys. Rev. D},
  volume = {81},
  issue = {3},
  pages = {034506},
  numpages = {15},
  year = {2010},
  month = {Feb},
  doi = {10.1103/PhysRevD.81.034506},
  url = {http://link.aps.org/doi/10.1103/PhysRevD.81.034506},
  publisher = {American Physical Society}
}


@article{PhysRevD.82.074501,
  title = {Scaling studies of QCD with the dynamical highly improved staggered quark action},
  author = {Bazavov, A. and Bernard, C. and DeTar, C. and Freeman, W. and Gottlieb, Steven and Heller, U. M. and Hetrick, J. E. and Laiho, J. and Levkova, L. and Oktay, M. and Osborn, J. and Sugar, R. L. and Toussaint, D. and Van de Water, R. S.},
  collaboration = {MILC Collaboration},
  journal = {Phys. Rev. D},
  volume = {82},
  issue = {7},
  pages = {074501},
  numpages = {15},
  year = {2010},
  month = {Oct},
  doi = {10.1103/PhysRevD.82.074501},
  url = {http://link.aps.org/doi/10.1103/PhysRevD.82.074501},
  publisher = {American Physical Society}
}


@article{PhysRevD.83.034503,
  title = {Tuning Fermilab heavy quarks in $2+1$ flavor lattice QCD with application to hyperfine splittings},
  author = {Bernard, C. and DeTar, C. and Di Pierro, M. and El-Khadra, A. X. and Evans, R. T. and Freeland, E. D. and G\'amiz, E. and Gottlieb, Steven and Heller, U. M. and Hetrick, J. E. and Kronfeld, A. S. and Laiho, J. and Levkova, L. and Mackenzie, P. B. and Simone, J. N. and Sugar, R. and Toussaint, D. and Van de Water, R. S.},
  collaboration = {Fermilab Lattice and MILC Collaborations},
  journal = {Phys. Rev. D},
  volume = {83},
  issue = {3},
  pages = {034503},
  numpages = {29},
  year = {2011},
  month = {Feb},
  doi = {10.1103/PhysRevD.83.034503},
  url = {http://link.aps.org/doi/10.1103/PhysRevD.83.034503},
  publisher = {American Physical Society}
}


@article{PhysRevD.67.014503,
  title = {Perturbative calculation of $O(a)$ improvement coefficients},
  author = {Harada, Junpei and Hashimoto, Shoji and Kronfeld, Andreas S. and Onogi, Tetsuya},
  journal = {Phys. Rev. D},
  volume = {67},
  issue = {1},
  pages = {014503},
  numpages = {8},
  year = {2003},
  month = {Jan},
  doi = {10.1103/PhysRevD.67.014503},
  url = {http://link.aps.org/doi/10.1103/PhysRevD.67.014503},
  publisher = {American Physical Society}
}


@article{Jpsi,
  title = {Precision tests of the $J/$\psi${}$ from full lattice QCD: Mass, leptonic width, and radiative decay rate to ${$\eta${}}_{c}$},
  author = {Donald, G. C. and Davies, C. T. H. and Dowdall, R. J. and Follana, E. and Hornbostel, K. and Koponen, J. and Lepage, G. P. and McNeile, C.},
  collaboration = {HPQCD Collaboration},
  journal = {Phys. Rev. D},
  volume = {86},
  issue = {9},
  pages = {094501},
  numpages = {21},
  year = {2012},
  month = {Nov},
  doi = {10.1103/PhysRevD.86.094501},
  url = {http://link.aps.org/doi/10.1103/PhysRevD.86.094501},
  publisher = {American Physical Society}
}


@article{Milc1,
  title = {Scaling studies of QCD with the dynamical highly improved staggered quark action},
  author = {Bazavov, A. and Bernard, C. and DeTar, C. and Freeman, W. and Gottlieb, Steven and Heller, U. M. and Hetrick, J. E. and Laiho, J. and Levkova, L. and Oktay, M. and Osborn, J. and Sugar, R. L. and Toussaint, D. and Van de Water, R. S.},
  collaboration = {MILC Collaboration},
  journal = {Phys. Rev. D},
  volume = {82},
  issue = {7},
  pages = {074501},
  numpages = {15},
  year = {2010},
  month = {Oct},
  doi = {10.1103/PhysRevD.82.074501},
  url = {http://link.aps.org/doi/10.1103/PhysRevD.82.074501},
  publisher = {American Physical Society}
}

@article{Davies,
      author         = "Davies, Christine",
      title          = "{Standard Model Heavy Flavor physics on the Lattice}",
      journal        = "PoS",
      volume         = "LATTICE2011",
      pages          = "019",
      year           = "2011",
      eprint         = "1203.3862",
      archivePrefix  = "arXiv",
      primaryClass   = "hep-lat",
      SLACcitation   = "
}

@article{El-Khadra,
      author         = "El-Khadra, Aida X. and Gamiz, Elvira and Kronfeld,
                        Andreas S. and Nobes, Matthew A.",
      title          = "{Perturbative matching of heavy-light currents at
                        one-loop}",
      journal        = "PoS",
      volume         = "LAT2007",
      pages          = "242",
      year           = "2007",
      eprint         = "0710.1437",
      archivePrefix  = "arXiv",
      primaryClass   = "hep-lat",
      reportNumber   = "FERMILAB-CONF-07-747-T",
      SLACcitation   = "
}

@article{DowdallVus,
      author         = "Dowdall, R.J. and Davies, C.T.H. and Lepage, G.P. and
                        McNeile, C.",
      collaboration = "HPQCD Collaboration",
      title          = "{Vus from pi and K decay constants in full lattice QCD
                        with physical u, d, s and c quarks}",
      journal        = "Phys.Rev.",
      volume         = "D88",
      pages          = "074504",
      doi            = "10.1103/PhysRevD.88.074504",
      year           = "2013",
      eprint         = "1303.1670",
      archivePrefix  = "arXiv",
      primaryClass   = "hep-lat",
      SLACcitation   = "
}

@article{RachelfB,
      author         = "Dowdall, R.J. and Davies, C.T.H. and Horgan, R.R. and
                        Lepage, G.P. and McNeile, C. and others",
      collaboration = "HPQCD Collaboration",
      title          = "{B, Bs, K and pi weak matrix elements with physical light
                        quarks}",
      year           = "2013",
      eprint         = "1309.4610",
      archivePrefix  = "arXiv",
      primaryClass   = "hep-lat",
      SLACcitation   = "
}

@article{McNeile,
      author         = "McNeile, C. and Davies, C.T.H. and Follana, E. and
                        Hornbostel, K. and Lepage, G.P.",
      title          = "{High-Precision $f_{B_s}$ and HQET from Relativistic
                        Lattice QCD}",
      journal        = "Phys.Rev.",
      volume         = "D85",
      pages          = "031503",
      doi            = "10.1103/PhysRevD.85.031503",
      year           = "2012",
      eprint         = "1110.4510",
      archivePrefix  = "arXiv",
      primaryClass   = "hep-lat",
      SLACcitation   = "
}

@article{Eduardo,
  title = {Highly improved staggered quarks on the lattice with applications to charm physics},
  author = {Follana, E. and Mason, Q. and Davies, C. and Hornbostel, K. and Lepage, G. P. and Shigemitsu, J. and Trottier, H. and Wong, K.},
  collaboration = {HPQCD Collaboration},
  journal = {Phys. Rev. D},
  volume = {75},
  issue = {5},
  pages = {054502},
  numpages = {23},
  year = {2007},
  month = {Mar},
  doi = {10.1103/PhysRevD.75.054502},
  url = {http://link.aps.org/doi/10.1103/PhysRevD.75.054502},
  publisher = {American Physical Society}
}

@article{MILCensemble,
  title = {Lattice QCD ensembles with four flavors of highly improved staggered quarks},
  author = {Bazavov, A. and Bernard, C. and Komijani, J. and DeTar, C. and Levkova, L. and Freeman, W. and Gottlieb, Steven and Zhou, Ran and Heller, U. M. and Hetrick, J. E. and Laiho, J. and Osborn, J. and Sugar, R. L. and Toussaint, D. and Van de Water, R. S.},
  collaboration = {MILC Collaboration},
  journal = {Phys. Rev. D},
  volume = {87},
  issue = {5},
  pages = {054505},
  numpages = {19},
  year = {2013},
  month = {Mar},
  doi = {10.1103/PhysRevD.87.054505},
  url = {http://link.aps.org/doi/10.1103/PhysRevD.87.054505},
  publisher = {American Physical Society}
}

@article{Rachel2,
      author         = "Dowdall, R.J. and Davies, C.T.H. and Horgan, R.R. and
                        Monahan, C.J. and Shigemitsu, J.",
      title          = "{B-meson decay constants from improved lattice NRQCD and
                        physical u, d, s and c sea quarks}",
      collaboration  = "HPQCD Collaboration",
      journal        = "Phys.Rev.Lett.",
      volume         = "110",
      pages          = "222003",
      doi            = "10.1103/PhysRevLett.110.222003",
      year           = "2013",
      eprint         = "1302.2644",
      archivePrefix  = "arXiv",
      primaryClass   = "hep-lat",
      SLACcitation   = "
}

@article{PDG,
  title = {Review of Particle Physics},
  author = {Beringer, J. and Arguin, J. -F. and Barnett, R. M. and Copic, K. and Dahl, O. and Groom, D. E. and Lin, C. -J. and Lys, J. and Murayama, H. and Wohl, C. G. and Yao, W. -M. and Zyla, P. A. and Amsler, C. and Antonelli, M. and Asner, D. M. and Baer, H. and Band, H. R. and Basaglia, T. and Bauer, C. W. and Beatty, J. J. and Belousov, V. I. and Bergren, E. and Bernardi, G. and Bertl, W. and Bethke, S. and Bichsel, H. and Biebel, O. and Blucher, E. and Blusk, S. and Brooijmans, G. and Buchmueller, O. and Cahn, R. N. and Carena, M. and Ceccucci, A. and Chakraborty, D. and Chen, M. -C. and Chivukula, R. S. and Cowan, G. and D'Ambrosio, G. and Damour, T. and de Florian, D. and de Gouv\^ea, A. and DeGrand, T. and de Jong, P. and Dissertori, G. and Dobrescu, B. and Doser, M. and Drees, M. and Edwards, D. A. and Eidelman, S. and Erler, J. and Ezhela, V. V. and Fetscher, W. and Fields, B. D. and Foster, B. and Gaisser, T. K. and Garren, L. and Gerber, H. -J. and Gerbier, G. and Gherghetta, T. and Golwala, S. and Goodman, M. and Grab, C. and Gritsan, A. V. and Grivaz, J. -F. and Gr\"unewald, M. and Gurtu, A. and Gutsche, T. and Haber, H. E. and Hagiwara, K. and Hagmann, C. and Hanhart, C. and Hashimoto, S. and Hayes, K. G. and Heffner, M. and Heltsley, B. and Hern\'andez-Rey, J. J. and Hikasa, K. and H\"ocker, A. and Holder, J. and Holtkamp, A. and Huston, J. and Jackson, J. D. and Johnson, K. F. and Junk, T. and Karlen, D. and Kirkby, D. and Klein, S. R. and Klempt, E. and Kowalewski, R. V. and Krauss, F. and Kreps, M. and Krusche, B. and Kuyanov, Yu. V. and Kwon, Y. and Lahav, O. and Laiho, J. and Langacker, P. and Liddle, A. and Ligeti, Z. and Liss, T. M. and Littenberg, L. and Lugovsky, K. S. and Lugovsky, S. B. and Mannel, T. and Manohar, A. V. and Marciano, W. J. and Martin, A. D. and Masoni, A. and Matthews, J. and Milstead, D. and Miquel, R. and M\"onig, K. and Moortgat, F. and Nakamura, K. and Narain, M. and Nason, P. and Navas, S. and Neubert, M. and Nevski, P. and Nir, Y. and Olive, K. A. and Pape, L. and Parsons, J. and Patrignani, C. and Peacock, J. A. and Petcov, S. T. and Piepke, A. and Pomarol, A. and Punzi, G. and Quadt, A. and Raby, S. and Raffelt, G. and Ratcliff, B. N. and Richardson, P. and Roesler, S. and Rolli, S. and Romaniouk, A. and Rosenberg, L. J. and Rosner, J. L. and Sachrajda, C. T. and Sakai, Y. and Salam, G. P. and Sarkar, S. and Sauli, F. and Schneider, O. and Scholberg, K. and Scott, D. and Seligman, W. G. and Shaevitz, M. H. and Sharpe, S. R. and Silari, M. and Sj\"ostrand, T. and Skands, P. and Smith, J. G. and Smoot, G. F. and Spanier, S. and Spieler, H. and Stahl, A. and Stanev, T. and Stone, S. L. and Sumiyoshi, T. and Syphers, M. J. and Takahashi, F. and Tanabashi, M. and Terning, J. and Titov, M. and Tkachenko, N. P. and T\"ornqvist, N. A. and Tovey, D. and Valencia, G. and van Bibber, K. and Venanzoni, G. and Vincter, M. G. and Vogel, P. and Vogt, A. and Walkowiak, W. and Walter, C. W. and Ward, D. R. and Watari, T. and Weiglein, G. and Weinberg, E. J. and Wiencke, L. R. and Wolfenstein, L. and Womersley, J. and Woody, C. L. and Workman, R. L. and Yamamoto, A. and Zeller, G. P. and Zenin, O. V. and Zhang, J. and Zhu, R. -Y. and Harper, G. and Lugovsky, V. S. and Schaffner, P.},
  collaboration = {Particle Data Group},
  journal = {Phys. Rev. D},
  volume = {86},
  issue = {1},
  pages = {010001},
  numpages = {1528},
  year = {2012},
  month = {Jul},
  doi = {10.1103/PhysRevD.86.010001},
  url = {http://link.aps.org/doi/10.1103/PhysRevD.86.010001},
  publisher = {American Physical Society}
}

\end{document}